\documentclass[aps,prb,letterpaper,twocolumn,nofootinbib,floatfix,showpacs]{revtex4}

\usepackage{graphicx}
\usepackage{epsfig}
\usepackage{notes2bib}
\usepackage{color}

\def\beq{\begin{equation}}
\def\eeq{\end{equation}}
\def\beqa{\begin{eqnarray}}
\def\eeqa{\end{eqnarray}}

\def\e{\epsilon}

\def\D{\Delta}
\def\del{\delta }
\def\e{\epsilon}
\def\cH{{\mathcal H}}

\def\Tr{\mathrm{Tr}}

\def\d{\mathrm{d}}

\def\del{\delta}

\def\avS{\langle S \rangle}

\begin{document}

\title{The effect of fluctuations - thermal and otherwise - on the temperature dependence of thermopower in aromatic chain single-molecule junctions}
\author{Yonatan Dubi$^{1}$}

\affiliation{$^1$Department of Chemistry and the Ilse Katz Institute for Nanoscale Science and Technology, Ben-Gurion University of the Negev, Beer-Sheva 84105, Israel}
\begin{abstract}
We report a theoretical study of the thermopower of single-molecule junctions, with focus on phenyl-based molecular junctions. In contrast to prior studies, thermal fluctuations of the torsional angle between 
the phenyl rings and variations in the position of the molecular level alignment with respect to the electrode Fermi energy were taken into account. Full thermopower histograms were obtained, and their 
dependence on the magnitude of the fluctuations was studied. We found that at large molecular orbital variations the thermopower becomes strongly dependent on the torsion angle and can even change sign. This 
results in a marked effect of fluctuations on the thermopower distribution, yielding an average  
thermopower at high temperatures that differs (smaller or larger) from the fluctuation-free value, depending on the strength of fluctuations. We therefore conclude that fluctuations should be taken into account both when extracting single-molecule 
parameters, such as the molecular level-Fermi level offset, and in predictions of the thermopower of molecular junctions.
  \end{abstract}
\maketitle

\section{Introduction} Current interest in the thermopower of single-molecule junctions is focused on two interrelated directions. The first is the expectation that molecular junctions may  
become the basis for high-efficiency thermoelectric devices, due to their low thermal conductivity and the large variety of possible junction compositions \cite{Nitzan2007,dubi2011colloquium}. The second is the notion that the thermoelectric response carries inherent information regarding the mechanisms dominating the electronic transport and the electronic 
structure of the junction (primarily the positions of the molecule's highest occupied (HO) and lowest unoccupied (LU) molecular orbitals (MO) with respect to the electrodes' Fermi level) \cite{Paulsson2003,Koch2004,Segal2005}. The relation between the two directions is 
simple - knowing the detailed electronic properties of the junction will help design junctions with better thermoelectric performance. Against this background, the last few years have 
seen huge advances in the measurement of thermopower in molecular junctions \cite{Reddy2007,Baheti2008,Malen2009,Malen2009a,Tan2011,Yee2011,Widawsky2012},
 accompanied by a large number of theoretical studies. \cite
{Murphy2008,dubi2008thermoelectric,Liu2009,Ke2009,Bergfield2009,Liu2009a,Wang2010,Leijnse2010,Sergueev2011,Liu2011,Liu2011a,Quek2011,Stadler2011,Nikolic2012,Burkle2012,Hsu2012,Balachandran2012}

The thermopower of a molecular junction (also known as the Seebeck coefficient) $S$ is defined as the (linear) voltage response to an applied temperature difference, $S=\lim_{\Delta T \rightarrow 0} -\Delta V/ \Delta T$. Most of the theoretical papers mentioned above have used the non-equilibrium Green's function (NEGF) approach \cite{Datta1997, DiVentra2008} in combination with density functional 
theory (DFT) to determine $S$. Within this framework, the Kohn-Sham (KS) equations for the hybrid metal-molecule-metal (finite) electrodes are solved self-consistently using various 
functionals \cite{Ke2009}. The KS orbitals are used to construct the Green's functions, which are, in turn, used to calculate the junction's transmission function, i.e.,  
the probability of an electron arriving from the electrode with energy $E$ to cross the junction. The self-energy that describes the electrodes is 
calculated either phenomenologically (using a simple level-broadening form \cite{Ke2009}) or by some self-consistent scheme \cite{Ke2004}. Once the transmission function 
$\tau(E)$ is calculated, it can be used to calculate the (temperature-dependent) thermopower $S$, as described in the following section. 

In recent years, the NEGF-DFT approach has been criticized, not only because of the approximation built into the use of KS orbitals to construct the Green's functions \cite
{DiVentra2008} or the use of different functionals \cite{Ke2009}, but because it cannot capture dynamical effects. Such effects may become very important for 
transport \cite{Varga2011,Myohanen2012,Evans2009}, and their role in determining the thermopower is currently not known. In addition, it has recently been shown that because of the 
inherent level broadening due to the self-energy, the NEGF approach "downplays" variations in the transmission function due to, e.g., the local density of states at the 
molecule-electrode point of contact. As a result, fluctuations of the energy offset $\D E$ between the HOMO and the electrodes' Fermi energy are overestimated \cite{Dubi2012}. 

The effect of fluctuations on the thermopower in molecular junctions has been a subject of recent theoretical studies 
\cite{Koch2004,Walczak2007,Galperin2008,Entin-Wohlman2010,Burkle2012,Tan2011,Pauly2011,Sergueev2011,Entin-Wohlman2012,Ren2012,Hsu2012}. However, neither the statistical nature (i.e. 
the full histogram of the thermopower) nor the effects of the variations in the molecular levels were discussed, even though all the experiments carry a 
significant statistical signature. In fact, even in the much more thoroughly studied field of electron transport in molecular junctions, the effect of various 
configurations on the statistical distribution were onlly recently discussed \cite{Reuter2012}. Addressing the effect of fluctuations theoretically is critical 
for properly analyzing experimental results 
and relating them to the electronic properties of the molecular junctions. 
We thus conducted a theoretical study of the effect of fluctuations on the temperature-dependence of the thermopower in model single-molecule junctions composed of biphenyl (two rings) and 
triphenyl (three rings) molecules attached to gold electrodes. 

Two sources of fluctuations were considered: (1) fluctuations in the energy offset $\D E$ between the Fermi energy of the 
electrodes and the HOMO level, and (2) thermal fluctuations of the torsion angle between the planes of the phenyl rings (the dihedral angle) \cite{Johansson2008,Burkle2012}. 
These two sources of fluctuations represent, respectively, fluctuations due to different reconstructions of the molecular junctions and thermal fluctuations within a given junction \cite{Malen2009,Guo2011}. Other sources of fluctuations, such as fluctuations in the molecule-electrode coupling may (and probably do) exist, and were discussed in past literature \cite{dubi2011colloquium}. However, here we focus on the above sources of fluctuations, which have been experimentally proven to exist\cite{Guo2011} and to bare importance of the thermopower\cite{Malen2009}. 

The main result reported here is that due to fluctuations, the average thermopower $S(T)$ can deviate substantially (higher or lower) from its bare, fluctuation free value. The full thermopower histogram is presented, and exhibits similarity to experimental results. The important conclusion that may be drawn from these findings is that in considerations of the thermopower of a molecular junction the full distribution of $S(T)$ must be used to extract information on the molecular junction. 

\section{Molecular junctions: model and calculation} 
In this work, we considered phenyl-based molecular junctions. Such junctions are prototypical molecular junctions whose conductance appears to be dominated by the torsion 
angle between the phenyl rings. \cite{Venkataraman2006,Mishchenko2010,Mishchenko2011} Note that the torsion angle can be adjusted by attaching to the rings different alkyl chains of different lengths 
\cite{Vonlanthen2009,Mishchenko2010,Mishchenko2011,Burkle2012}. The equilibrium torsion angle $\varphi_0$ is bistable at $\varphi_0\approx 30^o$ and $\varphi_0\approx 90^o$, with an energy barrier of about $\Delta \approx 0.3$ eV between them (when the molecule 
is in the junction) \cite{Sergueev2011}, \footnote{The energy barrier between the two stable states was calculated to be as be as small as $0.08$ eV for free biphenyl molecules \cite{Johansson2008}, but 
we take here the higher limit as a "worse-case". A smaller barrier would make the conclusions of this paper even stronger.} The thermopower of biphenyl molecular junctions was recently studied in detail 
with the NEGF-DFT method \cite{Sergueev2011, Burkle2012}. Sergueev {\sl et al.}\cite{Sergueev2011} studied the effects of inelastic (electron-phonon) scattering on the thermopower for the two stable torsion 
angles. Although they did take into consideration thermal fluctuations of the torsion angle, they investigated only a single anchor geometry between the molecule and the electrodes. Burkle {\sl et al.}\cite{Burkle2012} considered different 
anchor geometries and different end-groups as well as different angles, but did not include the temperature dependence of the fluctuations in their study. Neither of these groups took into account the fluctuations of the HOMO level or the full thermopower distribution. 

\begin{figure}[h]
\vskip 0.5truecm
\includegraphics[width=8.5truecm]{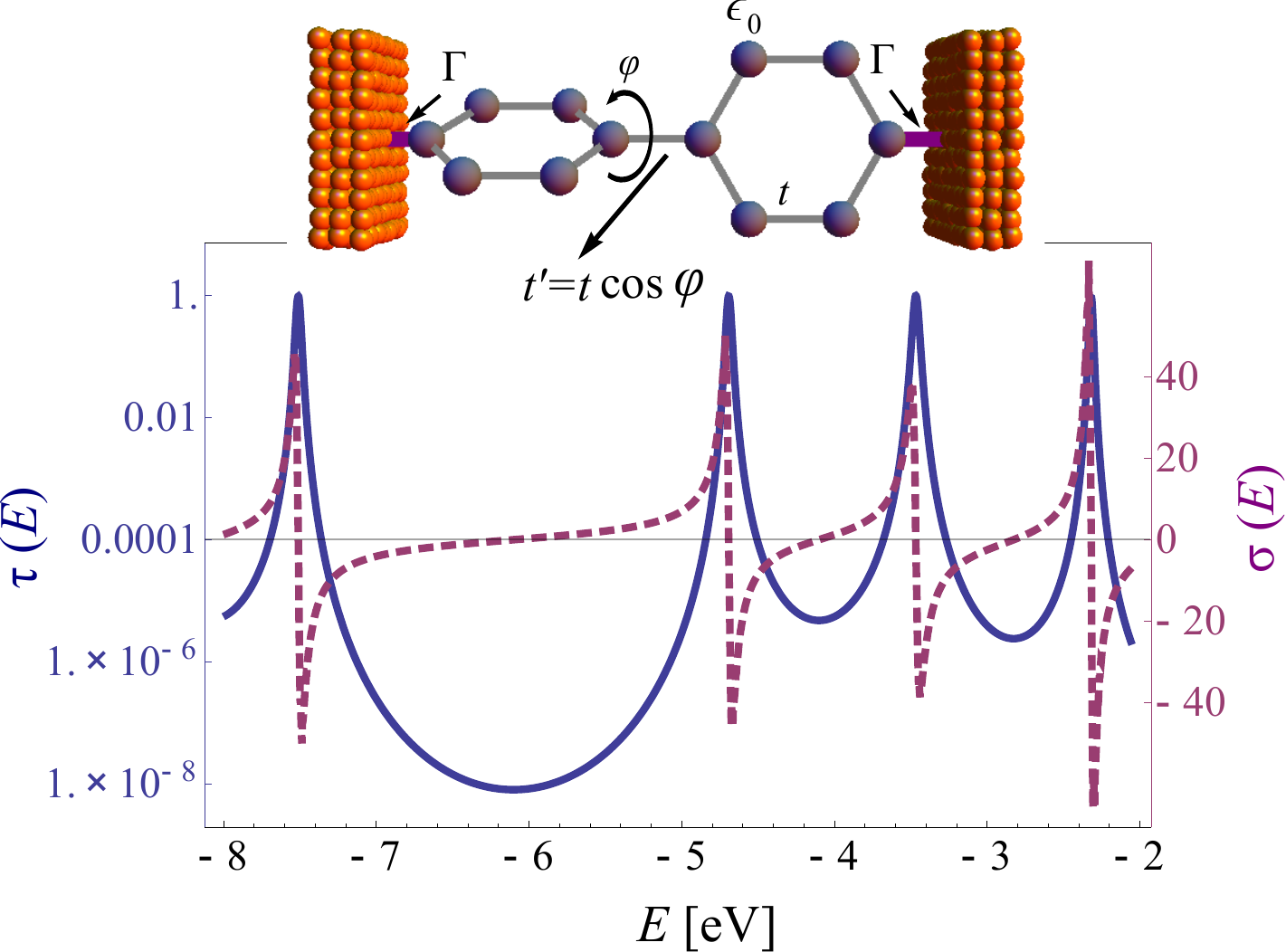}
\caption{Upper panel: Schematic representation of the molecular junction and the tight-binding parametrization. (a) Transmission function $\tau(E)$ (solid blue) and its logarithmic derivative $\sigma (E)$, which is proportional to the thermopower at low temperatures (dashed purple) as a function of energy $E$ for a CN-TT1 configuration biphenyl junction.  }
\label{fig1}
\end{figure}

To address the above-described deficiencies, we apply the $\pi$-orbital tight-binding (TB) model introduced in Ref.~\cite{Viljas2008} and parametrized by Burkle {\sl et al.}\cite{Burkle2012}. This model uses the transmission 
function from the full NEGF-DFT calculation as a fitting curve for a TB description of the biphenyl junction with different anchoring geometries. In the TB model, the molecule is described by a $\pi$
-orbital at every molecule location with an on-site energy $\e_0$ and a hopping term $t$ between orbitals belonging to the same phenyl ring. The inter-ring hopping is given by $t'=t \cos \varphi$, 
where $\varphi$ is the torsion angle. In the wide-band approximation, the self-energy of the electrodes is characterized by a level broadening $\Gamma /2$. The TB model is 
schematically depicted in the top panel of Fig.~\ref{fig1}, and the parameterization from Ref.~\cite{Burkle2012} for different end-groups is shown in Table \ref{table1}. We focus here on the TT1 (top) 
binding geometry, as the differences between the different geometries are rather minor.   

\begin{table}
\begin{center}
\begin{tabular}{c|ccc}
\hline \hline  & $\e_0$ & $t$ & $\Gamma$ \\ \hline \hline
S-TT1 & -4.0 & -1.9 & 0.96 \\ \hline
NH$_2$-TT1 & -4.30 & -2.29 & 0.6 \\ \hline
CN-TT1 & -6.10 & -2.0 & 0.14 \\ \hline
\end{tabular}
\caption[]{Parametrization of the TB model, following Burkle {\sl et al.} \cite{Burkle2012}  }
\label{table1}
\end{center}
\end{table}

Once the tight-binding Hamiltonian $\cH (\varphi)$ is parametrized (for a given torsion angle $\varphi$), we evaluate the retarded and advanced Green's functions $G^{r,a}(\e,\varphi)=\left(\e -\cH \mp i 
\eta \right)^{-1}$, from which the transmission function is calculated by $\tau(\e,\varphi)=\Tr \left[ \Gamma G^r(\e) \Gamma G^a(\e)\right]$. The thermopower can 
be calculated from $\tau(\e)$ by $S(T)=-L_1(T)/eT L_0(T)$, where $L_n(T)=\int dE \tau(E)(E-\mu)^n [-\partial f(E,T) / \partial E]$, $e$ 
is the absolute value of the electron charge, $f(E,T)=\left( \exp [(E-\mu)/k_B T]+1\right)^{-1}$ is the Fermi function, $k_B$ the Boltzmann constant, and $\mu$ is the chemical 
potential \cite{Paulsson2003,DiVentra2008,dubi2011colloquium}. At low temperatures, the expression for $S$ simplifies to $S(E)=-q(T) \sigma (E_F)$, where $q(T)=\pi^2 k^2_B T/3e$ and $\sigma (E)= \partial \log \tau (E) /\partial E$.  

In Fig.~\ref{fig1} we plot the transmission function $\tau(\e)$ (solid blue line) and the logarithmic derivative $\sigma(\e)$ (dashed purple line) as a function of 
energy for the CN-TT1 configuration, which has the smallest level broadening. The Fermi energy for Au electrodes is taken at $E_F \sim -5$ eV \cite{Burkle2012} (even though there is ambiguity of $\pm 0.1$ eV in this parameter \cite{Sergueev2011,Balachandran2012}).  The resonances at the HOMO and LUMO 
levels may be seen in the figure, with the resonance in $\sigma (E)$ close to them. 

\begin{figure}
\vskip 0.5truecm
\includegraphics[width=8.5truecm]{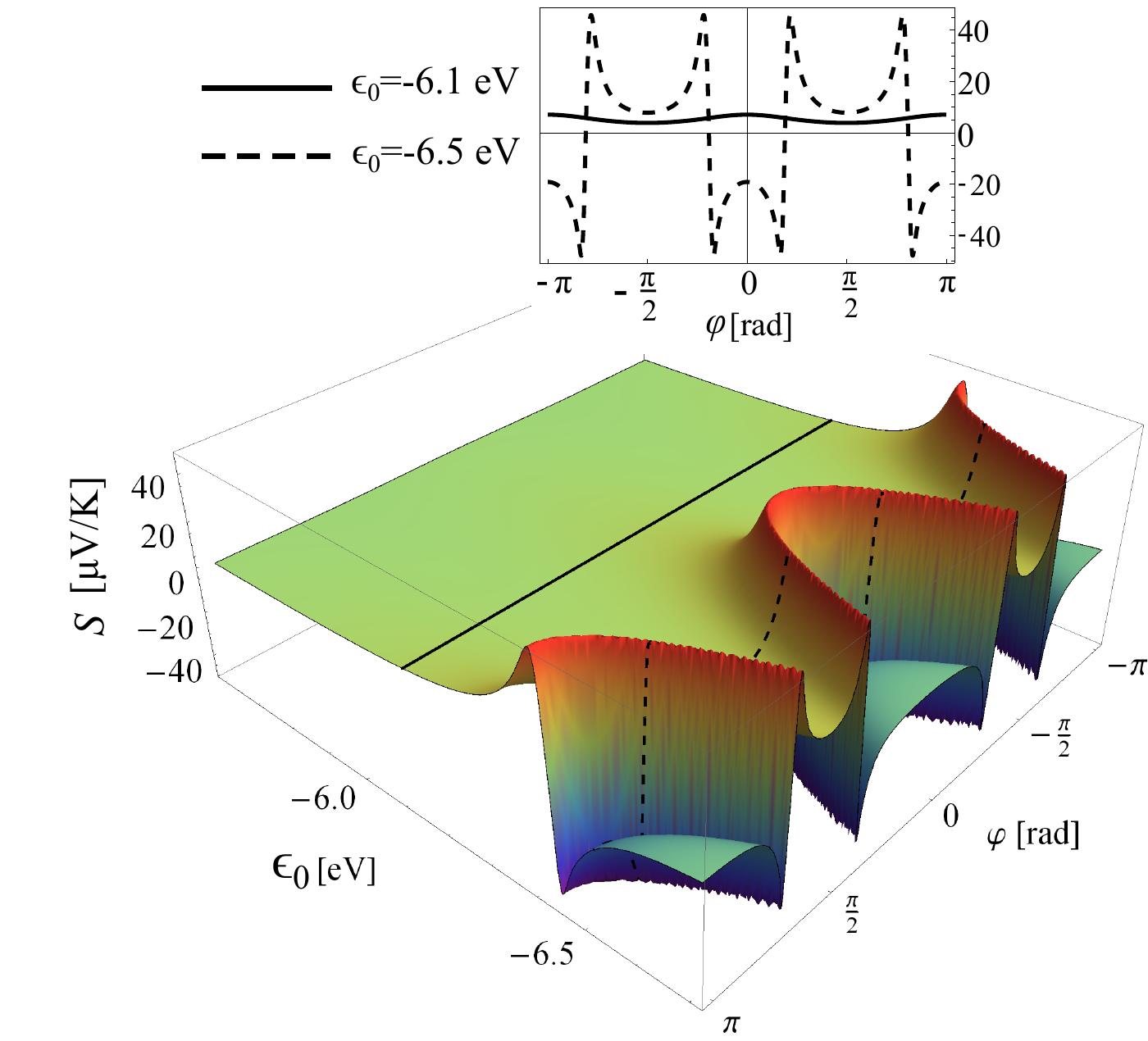}
\caption{Main panel: Thermopower $S$ of the CN-TT1 configuration as a function of molecular orbital energy $\e_0$ and biphenyl torsion angle $\varphi$. As $\e_0$ drops below $\sim -6.3$ eV, a strong dependence of $S$ on $\varphi$ is observed. Upper panel: A comparison of $S$ as a function of $\varphi$ for two values of orbital energy, $\e_0=-6.1$ and $ -6.5$ eV, demonstrating the dependence of $S$ on torsion angle. Note the possibility of sign change in $S$ as a function of angle.    }
\label{fig2}
\end{figure}

In Fig.~\ref{fig2} (left panel) a 3D-map of the thermopower (at $T=40$ K) is plotted as a function of energy $\e_0$, which determines the position of the HOMO level and the energy difference $\Delta E$ between the HOMO level and the Fermi energy (which is kept constant) and torsion angle $\varphi$ (measured from the equilibrium value $\varphi_0$) . The thermopower is calculated using the full integral form and not with the low-temperature form. For $\e_0 \sim -6.3$ eV, we observe 
an increasingly strong dependence of $S$ on the torsion angle. This strong dependence is evident from the upper panel of Fig.~\ref{fig2}, showing a plot of $S$ as a function of $\phi$ for two values of the energy offset, $\e_0=-6.1 $eV (solid line, 
the value taken from Ref.~\cite{Burkle2012}) and $\e_0=-6.5$ eV, for which the thermopower changes dramatically as a function of angle and even changes sign. The origin of this behavior can 
be understood from the observation that pushing $\e_0$ to lower values is equivalent to raising the Fermi level from its bare value. From Fig.~\ref{fig1} we see that $E_F=-5$ eV falls just below the 
resonance value at which the thermopower changes sign. Thus, pushing $\e_0$ down is equivalent to pushing the Fermi level closer to this resonance, where thermopower exhibits strong sensitivity to the torsion 
angle. 

\section{Fluctuation-averaged thermopower}
The strong sensitivity of $S$ to the torsion angle begins to evolve within $0.2$ eV of the bare value of $\e_0$. However, the value of $\e_0$ (or, more physically, the energy offset $\Delta E$ between 
the HOMO with respect to the Fermi energy) is known to vary substantially between different reconstructions of the same junction \cite{Malen2009}. Defining $\del E$ as the range of variations in $\Delta 
E$ (i.e., $\Delta E$ changes on a scale $2 \del E$ between realizations), we estimate $\del E$ to lie between a low of $\sim 0.5$ eV (extracted from transition voltage spectroscopy measurement\cite{Guo2011}) and a high of $\del E \sim 1.4$eV, as extracted from thermopower measurements \cite{Malen2009} (although this may be an 
overestimate \cite{Dubi2012}). Thus, taking into account  fluctuations of both $\e_0$ and $\varphi$ is essential in the calculation of the temperature dependence of the thermopower. 

To take both the variations in $\e_0$ and the thermal fluctuations of $\varphi$ into account, we calculate the average thermopower $\langle S(T) \rangle$ as a statistical average, taking $\e_0$ from a uniform 
distribution $U[\e_0^{(0)}\pm \del E]$, where $\e_0^{(0)}$ are the bare values taken from Table~\ref{table1}. It is reasonable to take a uniform distribution of $\e_0$, since the variations in $\e_0$ are due to junction reconstruction and are not thermal in origin. The phases are drawn from a thermal distribution $\mathcal{D}(\varphi) \propto \exp \left(-\frac{ \mathcal{E}(\varphi)}{k_B T}\right)$, where $\mathcal{E}(\varphi)$ is taken phenomenologically to be a cosine function with two minima at $\varphi_0=30^o $ and $ 90^o$, with an energy barrier $\Delta E=0.3$ eV between them. Formally, the average thermopower is thus:
\beq
\langle S(T) \rangle =\frac{ \int^{\e_0^{(0)}+\del E}_{\e_0^{(0)}-\del E} d \e_0 \int^{\pi}_0  \d\varphi S(T,\e_0,\varphi) \exp \left(-\frac{ \mathcal{E}(\varphi)}{k_B T}\right)}{2 \del E \int^{\pi}_0  \d\varphi \exp \left(-\frac{ \mathcal{E}(\varphi)}{k_B T}\right)}~.
\eeq

\begin{figure}
\vskip 0.5truecm
\includegraphics[width=8.5truecm]{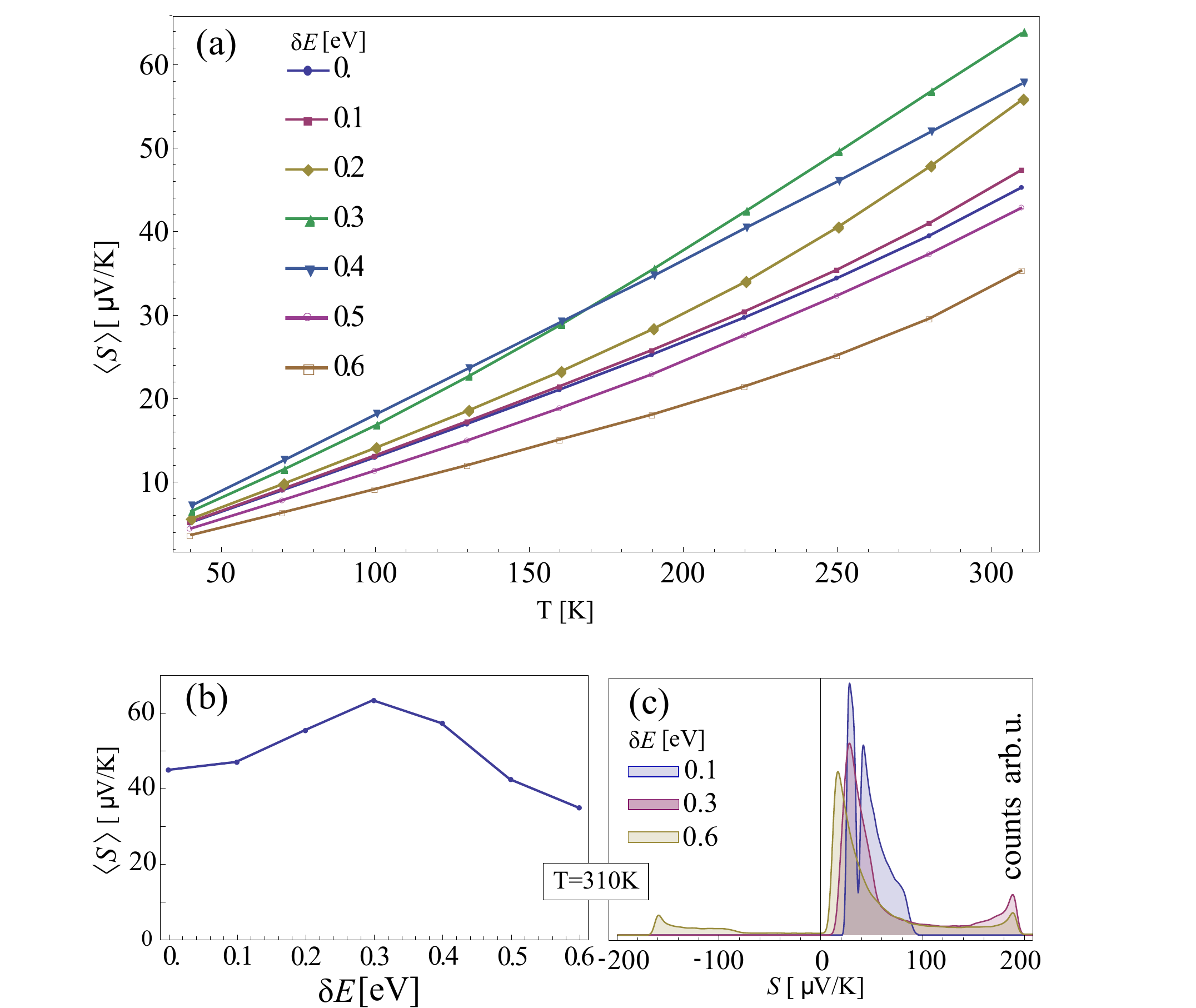}
\caption{(a) Average thermopower $\avS$ (averaged over torsion angles with a thermal distribution and molecular orbitals with a uniform distribution) as a function of temperature $T$ for different values of energy fluctuations range, $\del E=0,0.1,0.2,...,0.6$ eV. (b) $\avS$ at $T=310$ K as a function of $\del E$, demonstrating an inhomogeneous dependence. (c) Histograms of the thermopower $S$ for $\del E=0.1,0.3,0.6$ eV. Note the double-peak structure (which was similarly found in experimental observations) and the negative values at large $\del E$.}\label{fig3}
\end{figure}

In Fig.~\ref{fig3}(a), we plot the average thermopower $\avS$ as a function of the temperature, for the energy variation range $\del E=0,0.1,...,0.6$ eV (which are well within the experimentally relevant values \cite{Guo2011,Malen2009}), calculated for a CN-TT1 biphenyl molecular junction. The influence of the fluctuations can best be seen in Fig.~\ref{fig3}(b), where the average thermopower $\avS$ at $T=300$ K 
is plotted as a function of the energy variation range $\del E $. The average thermopower is not homogeneous in fluctuation strength, because as the fluctuations in $\e_0$ become large, there are more frequent realizations for which $\e_0$ is such that the Fermi energy is close to the transmission resonance, and thus 
the thermopower becomes strongly angle dependent (Fig.~\ref{fig2}). At a certain fluctuation strength, the thermopower can even change sign with the torsion angle (for certain 
realizations), giving rise to negative thermopower values, which then tend to reduce the average thermopower (hence the reduction at large $\del E$ and the 
non-monotonicity). This phenomenon can be seen in Fig.~\ref{fig3}(c), where the full thermopower $S$ histogram is plotted for  $\del E=0.1,0.3$ and $0.6$ eV. For $\del E=0.6$ eV, the histogram shows 
substantial weight on negative values of $S$. Note the histograms have a well-defined double-peak structure. Such structures, as well as sign-change of thermopower within the same junction, have been observed experimentally \cite{Reddy2007,Malen2009,Baheti2008}.

\begin{figure}
\vskip 0.5truecm
\includegraphics[width=6.5truecm]{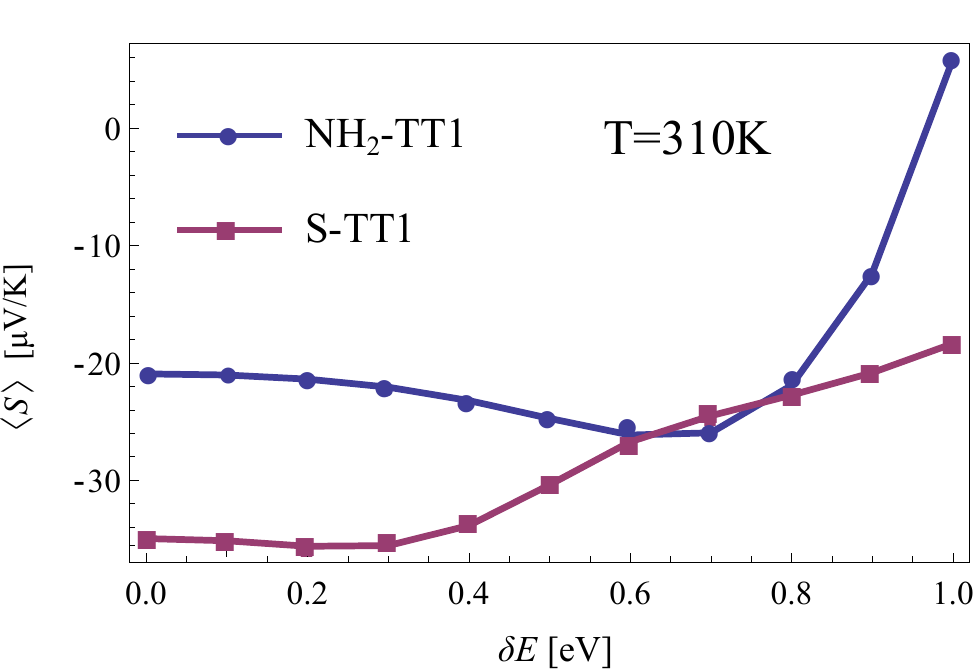}
\caption{Averaged thermopower $\avS$ as a function of orbital energy range $\del E$ at $T=310$ K for the NH$_2$-TT1 and S-TT1 configurations.}
\label{fig4}
\end{figure}

Out of all the configurations studied in Ref.~\cite{Burkle2012}, in the CN-TT1 configuration of the biphenyl junction the molecule is most weakly coupled to the electrodes, as manifested by the smallest of all values of $\Gamma$ (see Table ~\ref{table1}). Thus, the CN-TT1 configuration has the sharpest transmission resonance and 
thermopower line-shape and, as a result, the strongest dependence of the thermopower on the torsion angle close to the resonance (Fig.~\ref{fig2}). However, the deviation of the average thermopower from the fluctuation-free value due to fluctuations in $\e_0$ and $\varphi$ is not limited to junctions exhibiting sharp resonance. 
In Fig.~\ref{fig4} we plot the average thermopower $\avS$ at $T=310$ K as a function of $\del E$ for the NH$_2$-TT1 and S-TT1 configuration biphenyl molecular junctions, 
with $\Gamma=0.6$ and $0.96$ eV, respectively. Raising $\del E$ up to $1$ eV (corresponding to a variation in the HOMO-fermi energy of $2\del E=2$ eV), we find for these junctions a strong dependence of $\avS$ on $\del E$, including non-monotonic behavior and even a change of sign of $\avS$ at large $\del E$.

\begin{figure}
\vskip 0.5truecm
\includegraphics[width=8.5truecm]{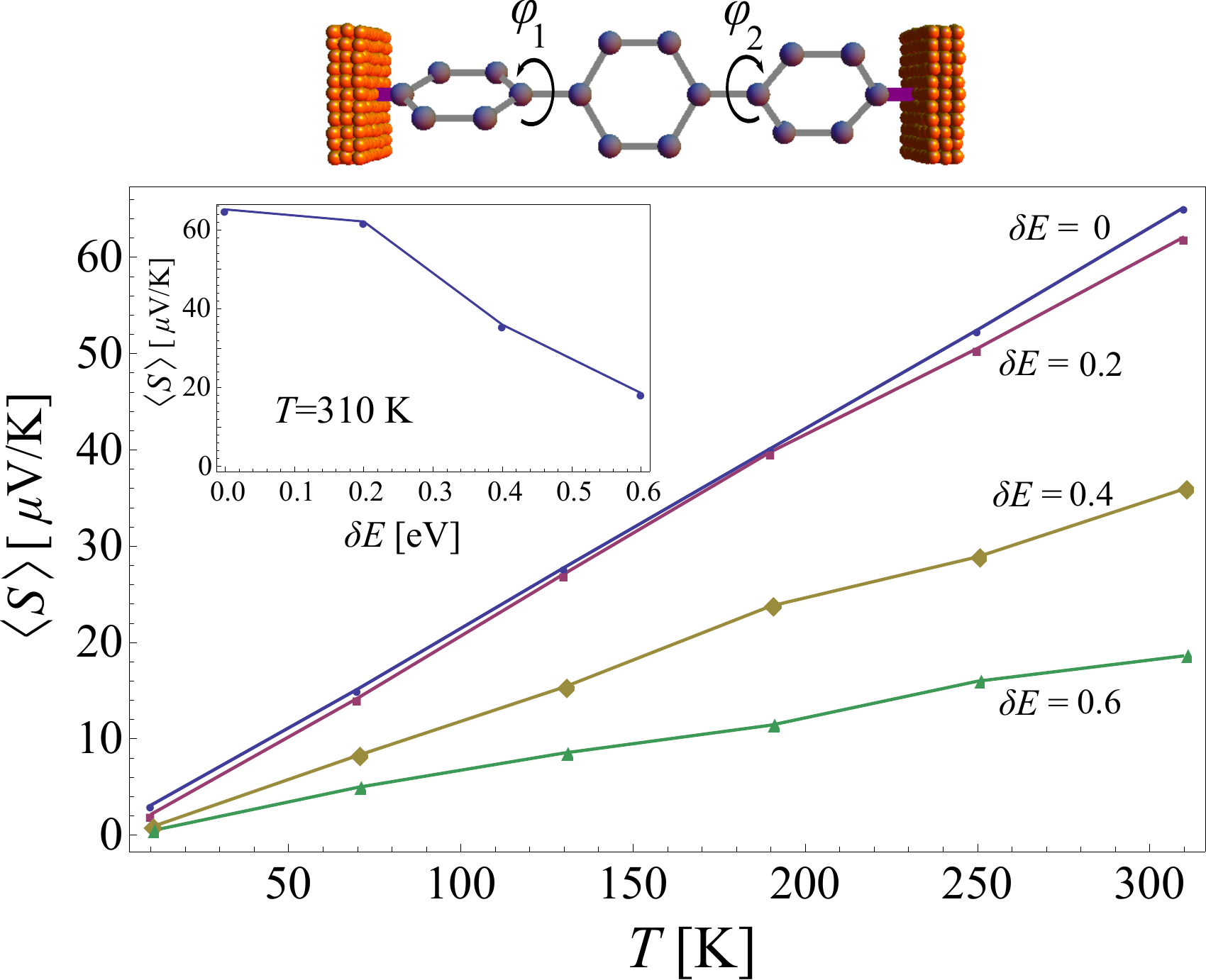}
\caption{Average thermopower $\avS$ as a function of temperature $T$ for different values of $\del E$ for the tri-phenyl molecular junction. 
Upper panel: Schematic representation of the tri-phenyl molecular junction with two torsion angles. Inset: Averaged thermopower $\avS$ as a function of orbital energy range $\del E$ at $T=310$ K. Note 
that for $\del E=0.6$ eV, the average thermopower is  $\sim 25 \%$ of its bare value, with no variations in $\e_0$ (i.e., $\del E=0$).}
\label{fig5}
\end{figure}

Finally, we calculate the temperature dependence of a triphenyl molecule in the CN-TT1 configuration, depicted in the upper panel of Fig.~\ref{fig5}. The molecular junction 
is now characterized by two torsion angles, $\varphi_1$ and $\varphi_2$. To calculate the transmission and from it the thermopower, we use the same TB 
parametrization as that of the biphenyl junction, under the assumption that the hopping element, the orbital level and the coupling to the electrodes should not depend (or very weakly depend) on the number of phenyl rings. We allow the two torsion angles $\varphi_1$ and $\varphi_2$ to thermally fluctuate 
around their equilibrium values, and the average $S(T,\e_0,\varphi_1,\varphi_2)$ is averaged over the fluctuations of $\e_0$ and the torsion angles. 

In Fig.~\ref{fig5}, $\avS$ is plotted as a function of temperature for an energy variation range of $\del E=0,0.2,...,0.6$ eV. Here we used the low-temperature form for the thermopower calculation. This is a good approximation, noting that the thermopower in Fig.~\ref{fig3} is almost linear in temperature, pointing that even room temperature is within the low-temperature regime (since the molecular energy scales are still much larger than the temperature).
 
A monotonic reduction of $\avS$ with $\del E$ is found, as seen 
in the inset, where we plot $\avS$ as a function of $\del E$  at $T=300$ K. Here we stress that for $\del E=0.6$ eV (a variation range of $\Delta E$ of $1.2$ eV), the value of $\avS$ 
is $S=17~\mu$V/K, which is $\sim 25 \%$ of its value without fluctuations. If, for instance, we were to use this value of $S$ to calculate the position of the HOMO level without taking fluctuations into account, the resulting HOMO level would deviate by $\sim0.4$ eV with respect to its real position. 
%\bibliography{achemso}

\section{Summary}
In summary, we calculated the thermopower of phenyl-based molecular junctions. Using a DFT-based TB parametrization \cite{Burkle2012}, we calculated the thermopower for four types of 
junctions: three biphenyl junctions with different end groups and a tri-phenyl junction with a CN end-group. As opposed to previous calculations, we calculated the full thermopower distribution taking 
into account both variations in the molecular orbitals (due to junction reconstruction) and thermal variations in the torsion angle between the phenyl-rings. 

Our calculations show that the thermopower histogram strongly 
depends on the magnitude of the fluctuations (characterized by the range of variations of the molecular orbital level $\del E$). The fluctuation-averaged thermopower can increase or decrease as a function 
of the magnitude of fluctuations, due to the appearance of realizations in which the molecular orbitals lie close to a transmission resonance. In these realizations, the thermopower changes sign with the phenyl 
ring torsion angle, leading to the appearance of additional peaks with a negative sign in the thermopower histogram and an overall reduction in the average thermopower. 

Our main message here is that fluctuations, both thermal and from other sources (e.g. HOMO position due to junction reconstruction), must be taken 
into account in calculations of thermopower of molecular junctions, if these calculations are to be compared to experimental data, or if the experimental data are to be used to extract relevant junction parameters. From 
the experimental side, since a major source of fluctuations is variations in the molecular level, it is highly desirable to find an experimental way to "pin" the molecular levels and reduce 
fluctuations. To this end, further research into the origin of these fluctuations should be carried out.

The author wishes to thank Dr. Y. Miller. This research was funded by a BGU start-up grant.

\end{document}